\documentclass[12pt,preprint]{aastex}
\usepackage{emulateapj5}
\begin{document}

\title{$Chandra$ ACIS-S Observations of Three Quasars with Low-Redshift Damped
Ly$\alpha$ Absorption: Constraints on the Cosmic Neutral-Gas-Phase Metallicity 
\\at Redshift $z\approx0.4$\altaffilmark{1}}

\author{David A. Turnshek\altaffilmark{2,3},
Sandhya M. Rao\altaffilmark{2,3}, \\
Andrew F. Ptak\altaffilmark{3,4},
Richard E. Griffiths\altaffilmark{3,5},
and Eric M. Monier\altaffilmark{3,6}}

\altaffiltext{1}{Based on data obtained with the NASA {\it Chandra X-Ray Observatory}.}

\altaffiltext{2}{Department of Physics \& Astronomy, University of Pittsburgh, Pittsburgh, PA 15260, USA}

\altaffiltext{3}{email: turnshek@quasar.phyast.pitt.edu, rao@everest.phyast.pitt.edu, 
ptak@skysrv.pha.jhu.edu, \newline griffith@astro.phys.cmu.edu, monier@astronomy.ohio-state.edu}

\altaffiltext{4}{Department of Physics \& Astronomy, Johns Hopkins University, 
Baltimore, MD 21218, USA}

\altaffiltext{5}{Department of Physics, Carnegie Mellon University, Pittsburgh, PA 15213, USA}

\altaffiltext{6}{Department of Astronomy, The Ohio State University, 
Columbus, OH 43204, USA}

\begin{abstract}

{\it Chandra X-Ray Observatory} ({\it CXO}) ACIS-S spectra of three
quasars which lie behind three foreground damped Ly$\alpha$ (DLA)
absorbers are analyzed in order to attempt to determine the amount
of photoelectric absorption due to metals present in their x-ray
spectra.  These absorbers are the three largest neutral hydrogen
column density absorption-line systems known at low redshift ($0.313
\le z_{abs} \le 0.524$).  They have HI column densities which lie in
the range $3\times10^{21} \lesssim N_{HI} \lesssim 5\times10^{21}$
atoms cm$^{-2}$.  At these redshifts the amount of photoelectric
absorption at x-ray energies is primarily an indicator of the
oxygen abundance.  Since the column densities of these systems are
so high, one would expect accurate metallicity measurements of them
to yield a robust estimate of the column-density-weighted cosmic
neutral-gas-phase metallicity at $z\approx0.4$.  We consider cases
where the DLA gas has solar element abundance ratios and ones with
the $\alpha$-group element abundance ratios enhanced. For the adopted
assumptions, the column-density-weighted cosmic neutral-gas-phase
metallicity of the non-enhanced elements (e.g. Zn) at $z\approx0.4$
likely lies in the range $\approx 0.04-0.38$ Z$_{\odot}$.

\end{abstract}




\section{INTRODUCTION}

Observations of the chemical evolution of the Universe provide
independent constraints on the processes which govern cosmic
nucleosynthesis and galaxy formation. This includes studies of the
chemical makeup of the Universe's gaseous and stellar components
with increasing redshift.  With regard to the gaseous structures,
it is widely recognized that intervening damped Ly$\alpha$ (DLA)
absorption-line systems in quasar spectra are important probes of
galaxy formation. They are not only the highest neutral hydrogen
column density systems, with $N_{HI} \ge 2\times10^{20}$ atoms
cm$^{-2}$, but consideration of them alone can be used to trace
nearly all of the {\it neutral gas in the Universe} out to at least
redshift $z=3.5$.  The work on DLA systems has included studies
of their cosmological incidence, mass density, and column density
distribution (e.g., see Rao \& Turnshek 2000, hereafter RT2000, and
references therein), the types and impact parameters of the luminous
galaxies associated with them (e.g., Le Brun et al. 1997, Turnshek et 
al. 2001, Nestor et al. 2002, Turnshek, Rao, \& Nestor 2002, Rao et al. 
2003) and their kinematic properties along the sight-lines to the background
quasars (e.g., Prochaska \& Wolfe 1999).  Of special interest to the
work presented here, Pettini and collaborators (Pettini et al. 1999
and references therein) have pioneered the method of using DLAs to
measure the cosmic neutral-gas-phase metallicity of the Universe.
However, it has not been possible to extend this work significantly
below $z\approx 0.6$ primarily due to the fact that, at these redshifts,
the measurements must be done from space.  This is because there are
no suitable unsaturated heavy-element resonance lines in the optical
which can be used to reliably measure metallicities in high-$N_{HI}$
systems. For example, ZnII$\lambda\lambda$2026,2062, which is a
good tracer of the neutral gas and is relatively unaffected by
depletion onto dust grains, is only visible in optical spectra
at redshifts $z>0.6$; and while CaII$\lambda\lambda$3934,3969
and NaI$\lambda\lambda$5891,5897 are visible in optical spectra,
they are affected by complicated depletion processes and possible
saturation effects.  Moreover, although many measurements at
$z=0$ can be obtained from observations within the Galaxy, these
measurements may not be representative of the Universe as a whole.
Therefore, given that neutral-gas-phase metallicities have been
measured for DLA systems at high redshifts, it would seem that the
more appropriate way to empirically establish the mean metallicity
of the Universe at low redshift is to measure the asymptotic trend
(and eventually the spread) in metallicity as one probes from high
redshifts to low redshifts.

Here we report attempts to use {\it Chandra X-Ray Observatory}
({\it CXO}) ACIS-S observations of three quasars to measure the amount
of photoelectric absorption due to metals in their spectra arising
from three foreground low-redshift DLA systems.  The three DLA
systems studied here have $0.313 \le z_{abs} \le 0.524$; they are
the largest HI gas columns presently known at low
redshifts (RT2000), with $3\times10^{21} \lesssim N_{HI} \lesssim
5\times10^{21}$ atoms cm$^{-2}$ as inferred from their DLA or 21
cm absorption lines. Since the K shell ionization edge of oxygen
occurs at a rest frame energy of 0.53 keV and oxygen has relatively
high absorptivity in comparison to the other heavy elements (Wilms,
Allen, \& McCray 2000), the amount of photoelectric absorption
in the observed 0.3-8 keV energy range, where {\it CXO} ACIS-S is
the most sensitive, is primarily an indicator of a DLA system's
oxygen abundance. Furthermore, because the three systems are
such high-$N_{HI}$ systems and the sum of their column densities
are similar to those probed at higher redshifts, definitive
measurements of the metal abundances in them would provide 
a robust determination of the cosmic neutral-gas-phase
metallicity at $z\approx0.4$.  This is an important constraint on
the process of cosmic chemical evolution;
specifically, it would be a measurement of the
metallicity of the neutral gas which is an important constituent of
matter in the cyclic process of galaxy formation (e.g., collapse of
gas to form large neutral gas structures, followed by the formation
of molecular clouds and star formation, eventually leading to metal
enrichment of the environment, and so on $-$ but see Allen 2002
for a different perspective).

In principle, the use of photoelectric absorption to measure DLA
metallicities has advantages. First, since atoms tied up in all
states (e.g., including ionized species, molecules, and grains)
contribute to photoelectric absorption, it can provide a more direct
measurement of the total metallicity.  Ionization corrections
and depletion onto grains need not be considered.  Second, due
to shielding below the hydrogen Lyman limit in the absorber rest
frame, the ionized hydrogen gas fraction is expected to be minimal
relative to the total column, and searches for molecular hydrogen
in DLA systems usually yield null results or very low molecular
hydrogen fractions (e.g., Levshakov et al. 2002, Petitjean et al.
2002, and references therein). Thus, the HI column densities
inferred from the DLA or 21 cm absorption lines of these systems
are indicative of the systems' total column densities. One would
therefore expect x-ray measurements of the DLA metallicities to
be straightforward, in the sense that it would not be necessary
to make corrections that depend on physical conditions. However,
the main limitation of our {\it CXO} ACIS-S observations is that they
are insufficient to readily resolve the oxygen k-shell ionization
edge at a particular redshift. Instead, any photoelectric absorption
signature is due to the cumulative effect of all the metals which
may be present in objects at all intervening redshifts, and it is
not possible to {\it definitively} determine the abundance of any
individual element, or even isolate the amount of absorption due
to metals at any particular redshift.

Our main aim here is to use {\it CXO} ACIS-S observations with
reasonable assumptions to derive a preliminary measurement of
the mean cosmic neutral-gas-phase metallicity of the Universe at
redshift $z\approx 0.4$. We consider the measurement preliminary for
three reasons.  First, since absorption edges are not specifically
identified, any detected absorption could arise from gas at some
other redshift. However, since the probability of encountering
a second strong x-ray absorber at $z>0$ along any of these
sight-lines is small, we will simply assume that any detected
absorption originates at the DLA redshifts. If this were not true,
our measurements would then represent upper limits on the DLA
systems' metallicities. Second, it is possible that the absorbers
we have studied here, while sampling the largest HI columns known
at low redshift, are biased against including more chemically
evolved systems which contain significant dust, thereby dimming
background quasars and eliminating them from current samples that
are being studied. This selection bias, if present, could cause
us to underestimate the mean cosmic neutral-gas-phase metallicity
at $z\approx 0.4$.  Third, {\it CXO} ACIS-S calibrations may change,
especially at energies $E<0.4$ keV.  Nevertheless, we believe that
the results reported here are valuable because they will, at the very
least, provide some insight on attacking this problem in the future.

The organization of this paper is as follows. In \S2 we present
the {\it CXO} ACIS-S quasar observations along the sight-lines through
three low-redshift DLA absorbers, and the analyses and assumptions
which lead to DLA metallicity determinations.  In \S3 we combine
our results with the DLA metallicity measurements of Pettini et
al. (1999) at $z>0.6$; this provides a preliminary estimate of the
evolution of the column-density-weighted cosmic neutral-gas-phase
metallicity at $z<3.5$. A discussion of the limitations of these
results and their implications is presented in \S4.

\section{Observations, Analyses, and Neutral-Gas-Phase 
Metallicity Determinations}

Observations of the three quasars were made with ACIS-S CCD \#7
on {\it CXO} (see {\it cxc.harvard.edu}). The journal of observations and 
other details are given in Table 1. First, we emphasize that an  
essential data processing step is to take into 
account the known decay of the ACIS quantum efficiency with 
time.\footnote{See 
{\it cxc.harvard.edu/cal/Links/Acis/acis/Cal\_prods/qeDeg/index.html}.}
Without this step, the derived metallicities for two of the DLA
absorbers would have been about 0.3 dex higher. 
The final processing of each
observation was performed using the {\it XAssist} software (Ptak
2001 and see {\it www.xassist.org}), which is a software
package that assists in x-ray data reprocessing, initial source
detection, and analysis.  The data reductions were performed using
CIAO 2.2 and the data were reprocessed using CALDB 2.9.  Briefly,
{\it XAssist} performs the basic data reduction steps recommended
by the {\it Chandra X-Ray Center} ({\it CXC}) ``threads.'' We also
implemented an optional step which removed the 0.5 pixel position
randomization. Sources were detected using the {\it wavdetect}
software, which detects sources using a wavelet procedure, and the
background light curves were examined, which permitted removal of
data taken during times of background flaring. Images of detected
sources were fitted with elliptical Gaussian models in order to
determine their spatial extent. The spectra were then binned to 20
counts per channel so that the $\chi^2$ statistic could be used to
examine the goodness of each spectrum fit.

The DLA absorption redshifts and HI column densities for the three
systems are included in Table 1.  The HI column densities for the
AO 0235+164 and PKS 1127$-$145 DLAs were derived by fitting Voigt
damping profiles to the Ly$\alpha$ absorption lines in their HST UV
spectra (RT2000).  Unfortunately, the HST UV spectrum of S4 0248+430
has very low signal-to-noise ratio, so the HI column density of
this system had to be derived from the its 21 cm absorption profile
(Lane \& Briggs 2001) for an assumed spin temperature of 700 K.
Adopting $T_s = 700$ K for this system is consistent with the
measured spin temperatures of other DLA absorbers (Carilli et
al. 1996; Lane et al. 1998).

Absorption of these quasars' x-ray spectra is due to hydrogen,
helium, and metals in the Milky Way Galaxy's ISM and in other
intervening systems.  To derive DLA metallicities we used the
publicly-available {\it XSPEC} software package (Arnaud 1996 and see
$xspec.gsfc.nasa.gov$), which is often used for analysis of
x-ray spectra.  We have limited our analysis to the $0.4<E<8$ keV region 
due to calibration uncertainties outside this range, especially below
0.4 keV.  We proceeded as follows.  We assumed that a power-law with
index $\Gamma$ fitted to each quasar's spectrum at energies $2<E<8$
keV could be extrapolated to lower energies to provide a reliable
indication of the {\it unabsorbed} quasar x-ray spectrum. This is
a reasonable assumption, since in no case is the absorbing column
large enough to affect the continuum at $E>2$ keV. Thus, the $2<E<8$ keV
range fixes both the slope and normalization of the x-ray spectrum. 
Our tests using
other quasars with similar emission redshifts in the {\it CXO} archives, 
which are not in environments where absorption at the emission 
redshift may be prevalent and which have no known DLA absorption, 
show that such extrapolations are
generally consistent with quasar spectra in the $0.4<E<2$ keV range
after accounting for Galactic absorption.  For the three quasars
considered here, the best-fit derived $\Gamma$s over this energy
range are reported in Table 1. We assumed that the most accurate
HI column densities of Galactic ISM gas along the sight-lines to
the three quasars are those derived from the
21 cm all-sky emission map of Hartmann \& Burton (1997).
These Galactic HI column densities are specified in Table 1.
However, it should be noted that the 0.5\degr\ $\times$ 0.5\degr\ angular
resolution of the HI map is much larger than the pencil beam of a
quasar line of sight. Therefore, the HI column density
derived from a 21 cm emission observation may only be approximate since
the measured value is an average over the size of the beam.  
We also assumed that the metallicity of the Galactic gas is given
by the ISM metallicity model of Wilms et al. (2000).  Wilms et
al. present a good justification for the metallicities they adopt.
The oxygen abundance in their model is 0.58 Z$_{\odot}$, while the
metallicities of nitrogen and carbon are 0.76 Z$_{\odot}$ and 0.60
Z$_{\odot}$, respectively.  The Wilms et al. ISM metallicity model is
one of the options incorporated into the {\it XSPEC} software package.
At the same time, in order to assess how our final results on DLA
metallicities depend on Galactic-gas metallicity, we also derived DLA
metallicities under the assumptions of zero and solar Galactic-gas
metallicity, where the solar metallicity is that specified by
Feldman (1992).  We note that recent investigations
generally indicate that {\it the Galactic-gas metallicity is
demonstrably sub-solar}, i.e., $\approx0.65$ Z$_{\odot}$ (e.g.,
Sembach et al. 1995, Roth \& Blades 1995, and Meyer, Jura, \&
Cardelli 1998).  For one Galactic sight-line, Meyer et al. (1994)
reported a Galactic-gas metallicity of 0.4 Z$_{\odot}$.

For the assumptions given above, and for two different cases of {\it
relative} DLA metal abundances, we then used {\it XSPEC} to determine
the best-fit overall metallicities for the three DLA systems.
For one case we assumed that the relative DLA metal abundances had
solar ratios as specified by Feldman (1992), and for the other case
we assumed that the $\alpha$-group element relative abundances
(specifically O, Ne, Mg, Si, S, Ar, and Ca) were enhanced by a
factor of 2.5 relative to the Feldman (1992) solar metal ratios.
Consideration of the enhanced $\alpha$-group element case seems
appropriate since the x-ray measurements should be most affected by
the oxygen abundance, whereas at higher redshift the most reliable
measurements of the overall DLA metallicity come from studies
of zinc which tracks iron, but show little or no evidence for
depletion onto grains.  The need to consider the possible effects
of $\alpha$-group abundance enhancements is clearly evident from
studies of a variety of stellar populations (e.g., see Prochaska
et al. 2000 and references therein).  The degree of $\alpha$-group
metals enhancement generally depends on the overall metallicity, with
[O/Fe]$=-0.4$[Fe/H] being close to the general trend. For example,
[Fe/H]$=-1$ (i.e., a metallicity of 0.1 Z$_{\odot}$) is indicative
of an enhancement of $\alpha$-group metals by a factor of 2.5 times
solar. Since the higher-redshift DLA measurements of Pettini et
al. (1999) indicate that zinc metallicities are typically $\lesssim
0.1$ Z$_{\odot}$, considering a case where $\alpha$-group element
abundances are enhanced by a factor of 2.5 seems appropriate.

The derived overall DLA metallicities and errors are reported in
Table 2 for the six combinations of assumptions about Galactic-gas
metal ratios (three different cases) and DLA-gas metal ratios
(two different cases) which were outlined above.  Also, because
the error source was quite different, the error in photon index
was not propagated to arrive at the quoted errors; but this is an
insignificant component to the error budget. This is especially true
given the various interpretive assumptions we have made (i.e., the
six combinations of assumptions mentioned above plus assumptions
put forth in \S1). Uncertainties in the Galactic and DLA HI column
densities have also not been propogated, but again, these 
contributions to the error budget are small relative to the 
interpretive assumptions. The $\chi^2$ statistic per number of degrees
of freedom is listed in Table 2 for each fit. This fit refers to
the entire $0.4<E<8$ keV range. These statistics show all of the
fits to be fairly reasonable, so this is an indication that x-ray
observations at such low resolution are unable to distinguish
between the six cases.  Because of the overwhelming evidence
that the Galactic-gas ISM metallicity is likely to be close to
that specified by Wilms et al. (2000), we adopt these results in
the remainder of the paper.

The left and right panels of Figures $1-3$ show the effect of DLA-gas
absorption on the quasar x-ray spectrum. Both panels include a
power-law fit in the $2-8$ keV range that is extrapolated  to 0.4 keV
to obtain the unabsorbed quasar x-ray spectrum and  absorption by
Galactic gas according to the Wilms et al. (2000) ISM metallicity
model. The right panel shows the best-fit model with DLA-gas
absorption also included. The results indicate that, for the AO
0235+164 and S4 0248+430 absorbers, the overall DLA metallicities may
be substantial, i.e., $> 0.1$ Z$_{\odot}$, and metallicities only a
factor of $\approx 2$ less than Galactic ISM values cannot be ruled
out.  In \S4.1.2 we do, however, discuss the possibility that another
significant absorber may be present in the S4 0248+430 spectrum. As
for the PKS 1127$-$145 DLA absorber, for our adopted assumptions we
find that there is no evidence for absorption {\it due to metals} at
the DLA redshift. In fact, our analysis {\it under-predicts} the x-ray
flux in the $0.4<E<2$ keV region even when the DLA gas metallicity is
zero. This can be seen by inspecting the right panel of Figure 3,
where the x-ray spectrum is seen to lie slightly above the model fit
which has zero DLA metallicity. This may indicate a problem with our
assumptions in the PKS 1127$-$145 analysis, which is further discussed
in \S4.1.3.
 
We note that Bechtold et al. (2001) also recently performed an
{\it XSPEC} analysis to study the metallicity of the DLA absorber in PKS
1127$-$145 (see both our Figure 3 and Bechtold et al.'s Figure 3).
However, pile up, which is a problem with the data (G. Chartas, {\it 
private communication}), was not considered in their analysis.
We attempted to model and remove this effect.
While the effects of pile up can be mitigated
by going off axis, one would have needed to be $>5$ arcmin off axis
to begin to obtain a significant reduction in pile up (G. Chartas, {\it
private communication}).

In the Bechtold et al. analysis, some emphasis was placed on exploring
the parameter spaces of x-ray spectrum fits, including the possibility
of absorption at other redshifts. Ignoring the fact  that pile
up was not considered in their analysis, their work indicated that,
indeed, the most plausible redshift for x-ray absorption was the DLA
redshift ($z_{abs}=0.313$).  For Galactic absorption along the PKS
1127$-$145 sight-line they used the observations of Murphy et
al. (1996) and adopted an HI column density of $3.8\times10^{20}$
atoms cm$^{-2}$, which is about 16\% larger than what we used based on
the  observations of Hartmann \& Burton (1997). The
metallicity they adopted for the Galactic ISM gas was presumably
solar. However, when making fits to any DLA x-ray absorption, they
chose to fix the DLA gas metallicity to be either zero (primordial) or
solar (processed),  and then they determined best-fit $N_{HI}$ values for
absorption of the x-ray spectrum.  They found that solar metallicity
gas with $N_{HI}=1.2\times10^{21}$ atoms cm$^{-2}$ provided the best
fit to the DLA-absorbed x-ray spectrum. Since this is 23\% of the HI
column indicated by the fit to the DLA line in the HST UV spectrum
(RT2000), they inferred a metallicity of 0.23 Z$_{\odot}$ for the PKS
1127$-145$ DLA absorber. However, they also showed that zero
metallicity  DLA gas at $z_{abs}=0.313$ with $N_{HI}=3.3\times10^{21}$
atoms cm$^{-2}$ can adequately absorb and explain the x-ray spectrum.
But in this case the required $N_{HI}$ value is
still only 63\% of the HI column derived from the HST spectrum.  This
conclusion is implausible since (as Bechtold et al. also note) metal
absorption lines {\it are} observed in the quasar spectrum at the DLA
redshift and the HI column density {\it is} greater than
$N_{HI}=3.3\times10^{21}$  atoms cm$^{-2}$. It would then follow that
DLA gas with the known column  density, i.e., $N_{HI}=5.1\times10^{21}$ 
atoms cm$^{-2}$,  cannot have a metallicity of 23\% solar and 
still be consistent with the x-ray spectrum.

The appropriate way to proceed, which is the approach we took in our
analysis, is to run the {\it XSPEC} software by fixing 
the DLA HI column density (since it is known) while allowing the
overall DLA absorber metallicity to be a free parameter.  This takes
advantage of the fact that fitting a Voigt profile to the UV DLA
absorption line provides a {\it metallicity-independent} method of
determining the HI column density.

\section{Column-Density-Weighted Cosmic Neutral-Gas-Phase Metallicity}

The integrated HI column densities of 42 high-redshift DLA systems
in five arbitrary redshift bins (e.g., see Pettini et al. 1999)
is comparable to the integrated HI column density of the three
systems we have studied here at $z\approx0.4$. This is illustrated
in Figure 4.  The distinguishing difference of the summed HI column
densities illustrated in Figure 4 is that the lowest redshift
bin (this work) pertains to metallicity measurements at x-ray
energies, while the higher redshift bins pertain to metallicity
measurements made at UV/optical wavelengths.  From this figure it
is clear that if metallicity determinations of sufficient accuracy
existed for the three individual low-redshift DLA systems studied
here, a very interesting result could be inferred on the mean
column-density-weighted cosmic neutral-gas-phase metallicity at
$z\approx0.4$. For this work, we adopt a definition analogous to
the one used by Pettini et al. (1999) to determine the mean 
column-density-weighted metallicity of Zn. In the absence 
of an effect which would bias our
sample (see \S1), such a measurement would represent the actual {\it
cosmic} neutral-gas-phase metallicity. Moreover, such a determination
would be robust in the sense that future metallicity measurements
of many lower-$N_{HI}$ systems would only slightly modify the result
because of their low weight.  

For the assumption of Galactic ISM gas with the metallicity
specified by Wilms et al. (2001) we have determined the following
from the {\it XSPEC} fits. If DLA gas has solar ratio metals,
the column-density-weighted neutral gas-phase metallicity at
$z\approx0.4$ is $0.19 \pm 0.11$ $Z_{\odot}$. Alternatively, if the
DLA gas has $\alpha$-group metals which are enhanced by a factor
of 2.5 over solar ratios, we find a metallicity of $0.10 \pm 0.06
$ $Z_{\odot}$ for the non-enhanced elements (e.g. Zn).  In addition, if one
were to take the view that the PKS 1127$-$145 DLA metallicity results
should be excluded from this analysis (see \S2 and \S4.1.3), the
results become $0.31 \pm 0.07$ $Z_{\odot}$ (for solar ratio metals)
and $0.15 \pm 0.04$ $Z_{\odot}$ (for non-enhanced metals mixed with
enhanced $\alpha$-group metals).

As discussed earlier, depletion onto grains is not a concern for x-ray
derived metallicities.  Thus, it is possible to make a comparison
between the x-ray derived low-redshift DLA metallicity results and
those derived from the ZnII lines in high-redshift DLAs.  This is
because Zn is not significantly depleted onto grains and the columns
of Zn not in the singly-ionized state are expected to be relatively
small, especially for the large HI column densities involved here
(e.g. Vladilo et al. 2001).  Since we quote metallicities for the
non-enhanced elements in both cases in the preceding paragraph,  the
results can be directly compared to the DLA [Zn/H] metallicities
considered by Pettini et al. (1999) at high redshift.  These
comparisons are made in Figure 5 (all three DLA absorbers) and Figure
6 (excluding the PKS 1127$-$145 DLA absorber).   The effect of using a
Galactic-gas metallicity other than the ISM values of Wilms et
al. (2001) can be judged from the range of results presented in Table
2.

Figures 5 and 6 show that the x-ray observations cannot be used to
rule out a significant increase in the column-density-weighted
neutral gas phase metallicity with decreasing redshift by
$z\approx0.4$.  However, given the various assumptions we have made,
and the formal  errors in our metallicity determinations for these
assumptions, the results should also not be interpreted as evidence
for an  increase in metallicity with decreasing redshift.

\section{Discussion}

\subsection{Comments on Individual DLA Systems}

Of course, analysis of the {\it Chandra} data is not limited to
results on DLA absorption in these quasars. Additional analysis on the
fields containing these quasars can be found in the {\it XAssist}
output (see {\it www.xassist.org}).  With regard to individual DLA
systems, various details should be noted here for each of the objects
studied. These are discussed below.

\subsubsection{AO 0235+164 ($z_{abs}=0.524$)}

AO 0235+164 is a well-known object that has properties of so-called
optically violent variables (OVVs).  It has been described as a
blazar, a BL Lac object, and a high-polarization QSO (e.g., see
Burbidge et al. 1996 and references therein). The high column density
absorption system at $z_{abs}=0.524$ was first discovered in
redshifted 21 cm absorption by Roberts et al. (1976).  Subsequently,
Wolfe et al. (1982) showed the 21 cm absorption to be variable and
Briggs (1983) proposed a kinematic model for the absorption.

Cohen et al. (1999) used their HST UV spectrum of AO 0235+164 to study
the DLA line associated with the 21 cm absorption and deduce the HI
column density and other properties of the absorber.  They found
$N_{HI}=5\times10^{21}$ atoms cm$^{-2}$; they also reported evidence
for dust at the absorption redshift, but at a significantly lower
level than what would be expected for a Galactic line-of-sight at
similar HI column density.  Our independent analysis of the HST UV
spectrum of AO 0235+164 yielded $N_{HI}=4.5\times10^{21}$ atoms
cm$^{-2}$ for the DLA absorber, which is the value adopted in our
analysis (Tables 1 and 2).

Using data taken with the Einstein Observatory Imaging Proportional
Counter in August 1980, Madejski (1994) performed an analysis which
showed the metallicity of the DLA absorber to be $2\pm1$ Z$_{\odot}$
for the assumptions he made. This is a higher metallicity than what we
infer from the superior $Chandra$ data.

Finally, HST imaging and follow-up spectroscopy of the field by
Burbidge et al. (1996) has led to a rather surprising serendipitous
discovery about the environment surrounding the AO 0235+164 DLA
absorber. There are two optical identifications of objects (called
``A1'' and ``A'') at redshift $z\approx0.524$ that lie within 2 arcsec
of the AO 0235+164 sight-line. The nearer object, A1, is separated by
1.08 arcsec from the sight-line and has a WFPC2 instrumental magnitude
of m$_{F702W}=21.3$.  It is a resolved emission-line galaxy, appearing
as a structure which is elongated 2 arcsec, with a brighter inner
region that has an extent of 0.5 arcsec.  Object A is separated by
1.96 arcsec from the sight-line and has m$_{F702W}=20.6$.  It appears
as an unresolved point source surrounded by faint nebulosity. Its
spectrum exhibits broad emission lines as well as broad absorption
lines (BALs), which have a maximum outflow velocity ($6000$ km
s$^{-1}$); this maximum velocity is on the low side of what is
normally observed in BAL QSOs. However, Burbidge et al. (1996) note
that, based on object A's magnitude and the surrounding faint
nebulosity, it might best be classified as a Seyfert 1 AGN, in which
case the outflow velocities are rather large in comparison to those
sometimes seen in AGN.  In any case, the presence of both an
emission-line galaxy (object A1) and a BAL QSO/AGN (object A) in the
vicinity of the AO 0235+164 DLA absorber is suggestive of a
chemically-evolved environment.

\subsubsection{S4 0248+430 ($z_{abs}=0.394$)}

The identification of the DLA absorber in S4 0248+430 is based on 21
cm absorption (Lane \& Briggs 2001).  As such, a DLA absorption line
was not used to derive its HI column density.  Instead, a reasonable
value of the spin temperature was assumed for the HI column density
derivation (\S2).  The fact that the continuum was too faint to
observe the DLA line in S4 0248+430's HST UV spectrum (RT2000) may be
an indication that significant dust is present along the sight-line.

This sight-line also represents a case where significant metal-line
absorption from objects other than the DLA system are clearly
present. However, it is unclear if  the column densities of these
other absorption systems are sufficient to give rise to x-ray
absorption.  In particular, in addition to the DLA system at
$z_{abs}=0.394$, Womble et al. (1990) identify two other distinct
low-ionization systems at $z_{abs}=0.052$ and $z_{abs}=0.452$.  The
rest equivalent widths of the MgII lines at $z_{abs}=0.452$ (Womble et
al.  1990) are small enough that this system is not likely to be in
the DLA regime (RT2000). Indeed, our HST spectrum of the Ly$\alpha$
line at $z_{abs}=0.452$ (RT2000) indicates that this system is not
damped. Thus, we conclude that the $z_{abs}=0.452$ does not give rise
to any appreciable x-ray absorption. However, as discussed below,
x-ray absorption from the $z_{abs}=0.052$ is a possibility.

The system at $z_{abs}=0.052$ is likely associated with a luminous
spiral galaxy at redshift $z=0.052$ which is offset 14.7 arcsec from
the sight-line to S4 0248+430 (Womble et al. 1990; Sargent and Steidel
1990).  At such low redshift, light from this extended luminous galaxy
overlaps and contaminates the image of the background quasar. Both
Womble et al. (1990) and Sargent \& Steidel (1990) detected CaII and
NaI absorption at this redshift in the quasar's spectrum. The
high-resolution spectrum of Womble et al. shows two primary absorption
components at this redshift. Their curve-of-growth analysis indicates
that the total Ca$^+$ column density of both components is
$\approx5.3\times10^{13}$ atoms cm$^{-2}$.  If there were no calcium
depletion onto grains and this were indicative of a neutral gas column
with solar abundances, the total HI column density would be
$\approx2.3\times10^{19}$ atoms cm$^{-2}$, which would not be large
enough to cause significant x-ray absorption. However, the
curve-of-growth error analysis of Womble et al. admits the possibility
that the Ca$^+$ column density is significantly higher. Moreover,
calcium is easily depleted onto grains and, as discussed above, there
is evidence for dust reddening along this sight-line.  Thus, we cannot
rule out the possibility that some of the x-ray absorption seen in the
S4 0248+430 spectrum comes from this system.

\subsubsection{PKS 1127$-$145 ($z_{abs}=0.313$)}

The properties of the various emission-line galaxies near redshift
$z=0.313$ along the sight-line towards PKS 1127$-$145 are described by
Lane et al. (1998), Nestor et al. (2002), and Rao et al. (2003).  Most
notably, there is a large patchy/irregular low surface brightness
structure which is very extended in the north-south direction and
$\approx3.5$ arcsec to the west of the sight-line.

As discussed in \S2, Bechtold et al. (2001) originally studied the
possibility of x-ray absorption from this DLA system.  This same group
has also made a detailed study of the {\it Chandra} x-ray images of
the field (Siemiginowska et al. 2002).  Recall that the main problem
we encountered in our analysis of x-ray absorption from this DLA
absorber (\S2) was that, even with zero metallicity DLA gas, after
Galactic absorption was taken into account the DLA absorption column
required to explain the x-ray spectrum at energies below 2 keV was
less than what was measured in the RT2000 HST UV spectrum. This
suggests that extrapolation of the power-law fit to the $2-8$ keV
region to lower energies under-predicts the x-ray flux in the $0.4-2$
keV region, indicating that there may be, for example, a significant
soft x-ray energy excess in PKS 1127$-$145.  Siemiginowska et
al. (2002) present some evidence for this; however, the flux from an
x-ray jet identified by Siemiginowska et al. in the x-ray image
appears to be negligible. In any case, estimating the metallicity of
this object is problematic.  We can only conclude that in this DLA
system there is presently no evidence for relatively high metallicity,
unlike the evidence in the other two DLA absorbers.

\subsection{Limitations of the Results and Implications}

The details of the objects studied (\S4.1) and the spectral resolution
of the {\it CXO} ACIS-S observations present limitations to the direct
detection of metals at the DLA redshifts.  Thus, our results and
interpretation depend on the reliability of the assumptions we have
made. For now, we believe that our assumptions represent the most
probable case.  This leads to the conclusion that we can not rule out
the possibility of significant evolution of the {\it cosmic}
neutral-gas-phase metallicity in the form of increasing metallicity
with decreasing redshift. In fact, Figure 6 suggests this trend, but
the results are also consistent with no evolution (Figure 5).

The statistics of low-redshift DLAs (RT2000) support the assumption
that the identified DLA system in each quasar spectrum is the dominant
large column density system along the sight-line to the quasar.
Except in the case of S4 0248+430, the possibility of other
significant absorption can be ruled out with some degree of certainty
(\S4.1).  Also, because our sample is a radio-selected one, it
probably does not suffer from the bias mentioned in \S1, namely that
chemically evolved DLA systems with significant dust are excluded from
the sample due to optical dimming of background quasars.  If present,
this selection bias would cause an underestimate of the mean cosmic
neutral-gas-phase metallicity at $z\approx 0.4$.

The results reported here show that {\it CXO} ACIS-S observations can
be used to investigate some important issues involving the x-ray
absorption properties of DLA gas. However,  our present results on the
objects considered here indicate  that x-ray grating observations are
required in order to definitively make DLA metallicity measurements at
x-ray energies. Such observations would not only permit individual
features from redshifted metals to be observed, it would also
eliminate the  simplifying assumption that the unabsorbed x-ray
spectrum can be modeled with a simple power law.

\bigskip

\centerline{\bf Acknowledgments}

\smallskip

We want to thank those involved with the {\it CXO} for making it a
user-friendly facility. In addition, we want to especially thank the
referee, Dr. George Chartas, for an excellent report which allowed us
to substantially improve the paper.


\clearpage

\begin{center}
\begin{deluxetable}{rcccccccrrl}
\rotate 
\tablewidth{0pc} 
\setlength{\tabcolsep}{0.5mm} 
\tablecaption{Journal of {\it Chandra X-Ray Observatory} ACIS-S
Observations} 
\tablehead{  
\colhead{} &  
\colhead{} &
\colhead{$N_{HI}^{GAL}$} & 
\colhead{DLA} &
\colhead{$N_{HI}^{DLA}$} & 
\colhead{Redshifted} & 
\colhead{Obs.} &
\colhead{Cleaned} & 
\colhead{Total} & 
\colhead{$2-8$ keV} &
\colhead{Program}\\[.2ex] 
\colhead{Quasar} &  
\colhead{$z_{em}$} &
\colhead{(10$^{21}$ cm$^{-2}$)$^1$} &
\colhead{$z_{abs}$} &
\colhead{(10$^{21}$ cm$^{-2}$)} & 
\colhead{DLA Oxygen} &
\colhead{Date} & 
\colhead{Exp Time} & 
\colhead{Counts} &
\colhead{$\Gamma^2$} & 
\colhead{ID (PI)}\\[.2ex] 
\colhead{} &
\colhead{} & 
\colhead{} & 
\colhead{} & 
\colhead{} & 
\colhead{K edge (keV)} & 
\colhead{} & 
\colhead{(ksec)} & 
\colhead{} & 
\colhead{} &
\colhead{}\\[.2ex] }

\startdata    AO 0235+164   &0.94&0.87&0.524&4.5$\pm$0.4&0.35&20/09/00&27.5& 3,577 &1.61$\pm$0.10&884 (Turnshek) \\ 
S4 0248+430   &1.31&0.89&0.394&3.9$\pm$0.6$^3$&0.38&27/09/00&22.8& 2,570&1.60$\pm$0.12&885 (Turnshek) \\ 
PKS 1127$-$145&1.19&0.33&0.313&5.1$\pm$0.9    &0.40&28/05/00&27.3&15,818&1.20$\pm$0.05&866 (Bechtold) \\ 
\enddata 
\tablecomments{(1) Galactic HI column density derived from the 21 cm emission map of Hartmann \&
Burton (1997).  Uncertainties are on the order of 2\%.
(2) The quoted error is approximately a 1$\sigma$
error derived from the $\chi^2$ statistics of the fit.  (3) The HI
column density is derived using the 21 cm data reported in Lane \&
Briggs (2001) and is slightly revised from the Lane  (private
communication) result reported in RT2000.}
\end{deluxetable}
\end{center}

\vskip 0.5in
\begin{deluxetable}{rccccccccl}
\scriptsize
\rotate
\tablewidth{0pc}
\setlength{\tabcolsep}{0.5mm}
\tablecaption{Derived Neutral-Gas-Phase Metal Abundances for Three Low-z DLAs$^1$}
\tablehead{ 
\colhead{} & 
\colhead{$N_{HI}^{GAL}$} &
\colhead{Galactic} &
\colhead{} &
\colhead{$N_{HI}^{DLA}$} &
\colhead{Best-Fit DLA} &
\colhead{Best-Fit DLA} &
\colhead{$\chi^2$ per dof $^4$} &
\colhead{$\chi^2$ per dof $^4$} \\[.2ex]
\colhead{Quasar} & 
\colhead{(10$^{21}$ atoms} &
\colhead{Metallicity} & 
\colhead{$z_{abs}$} &
\colhead{(10$^{21}$ atoms} &
\colhead{Metallicity (Z$_{\odot}$)} &
\colhead{Metallicity (Z$_{\odot}$)} &
\colhead{(solar ratio} &
\colhead{(enhanced $\alpha$} \\[.2ex]
\colhead{} & 
\colhead{per cm$^2$)} &
\colhead{Model$^2$} & 
\colhead{} &
\colhead{per cm$^2$)} &
\colhead{(solar ratio metals)$^3$} &
\colhead{(enhanced $\alpha$ metals)$^3$} &
\colhead{metals)} &
\colhead{metals)} \\[.2ex]
}
\startdata   
AO 0235+164   &0.87&0    &0.524&4.5$\pm$0.4&0.61$\pm$0.07&0.29$^{+0.04}_{-0.03}$&150/131&149/131 \\
              &    &ISM  &     &           &0.37$\pm$0.06&0.19$\pm$0.03&143/131&143/131 \\
              &    &Solar&     &           &0.24$\pm$0.06&0.11$\pm$0.03&142/131&142/131 \\

S4 0248+430   &0.89&0    &0.394&3.9$\pm$0.6&0.46$\pm$0.08&0.22$\pm$0.04& 86/100& 86/100 \\
              &    &ISM  &     &           &0.23$\pm$0.07&0.11$\pm$0.04& 84/100& 84/100 \\
              &    &Solar&     &           &0.10$^{+0.08}_{-0.07}$&0.04$^{+0.04}_{-0.03}$& 83/100& 83/100 \\

PKS 1127$-$145$^5$&0.33&0    &0.313&5.1$\pm$0.9&0 ($<0.0035$) &0 ($<0.0024$) &389/300&389/300 \\
                  &    &ISM  &     &           &0 ($<0.0032$) &0 ($<0.0022$) &367/300&367/300 \\
                  &    &Solar&     &           &0 ($<0.0029$) &0 ($<0.0020$) &475/300&371/300 \\
\enddata
 
\tablecomments{(1) The derived DLA metallicities hold for the
specified values of Galactic ($z=0$) $N_{HI}$ and metallicity. If
additional absorption is present, these results represent upper
limits on DLA metallicity (see text for assumptions and discussion).
(2) Three Galactic-gas metallicity models were considered: zero metallicity,
the ISM metallicity model of Wilms et al.  (2001), and the solar
metallicity model of Feldman (1992).  The ISM metallicity model used to
construct Figure 5 has an oxygen abundance of 0.58 Z$_{\odot}$, a nitrogen
abundance of 0.76 Z$_{\odot}$, and a carbon abundance of 0.60
Z$_{\odot}$ (see \S3). (3) The
DLA metallicities quoted here were inferred from the 
{\it XSPEC} results so that they could be compared to the zinc metallicity
results at high redshift (e.g., Pettini et al. 1999), where the
DLA zinc abundance is given relative to the solar zinc abundance.
In the case of DLA gas with $\alpha$-group elements enhanced by a factor
of 2.5, one sees that the inferred equivalent zinc metallicity is
lower; this is because the abundance of elements like oxygen are enhanced 
in the gas, allowing the abundance of elements like zinc to be lower and
yet produce the best fit to the x-ray spectrum.  The quoted errors
are approximately 1$\sigma$ for the assumptions made. (4) 
The $\chi^2$ per degree of freedom for the fit to the x-ray spectrum.
(5) Spectrum corrected for pile up. However, the inferred
low metallicity may be an artifact of assuming that the unabsorbed
x-ray spectrum can be described by a single power law.}
\end{deluxetable}

\clearpage

\begin{figure}
\plottwo{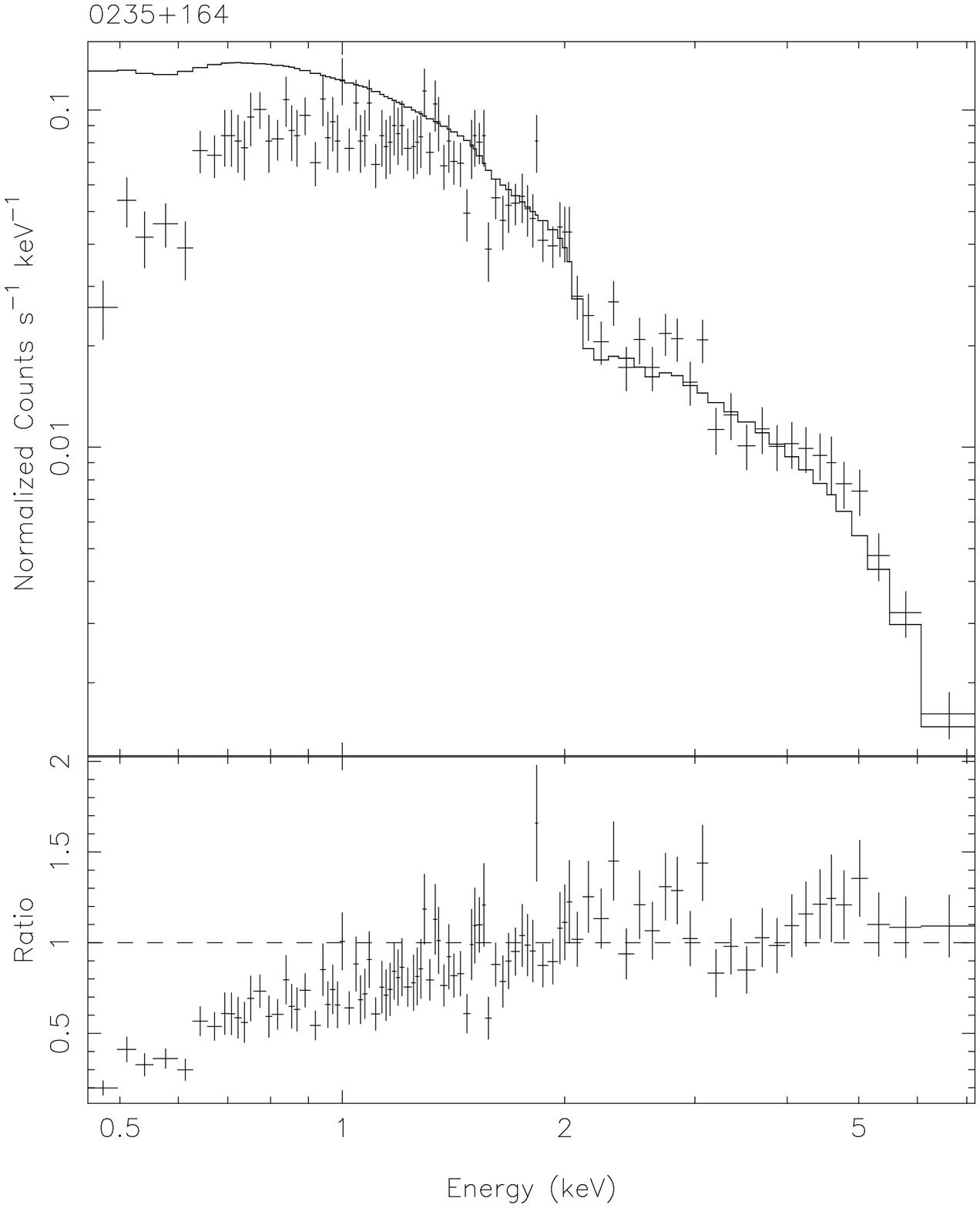}{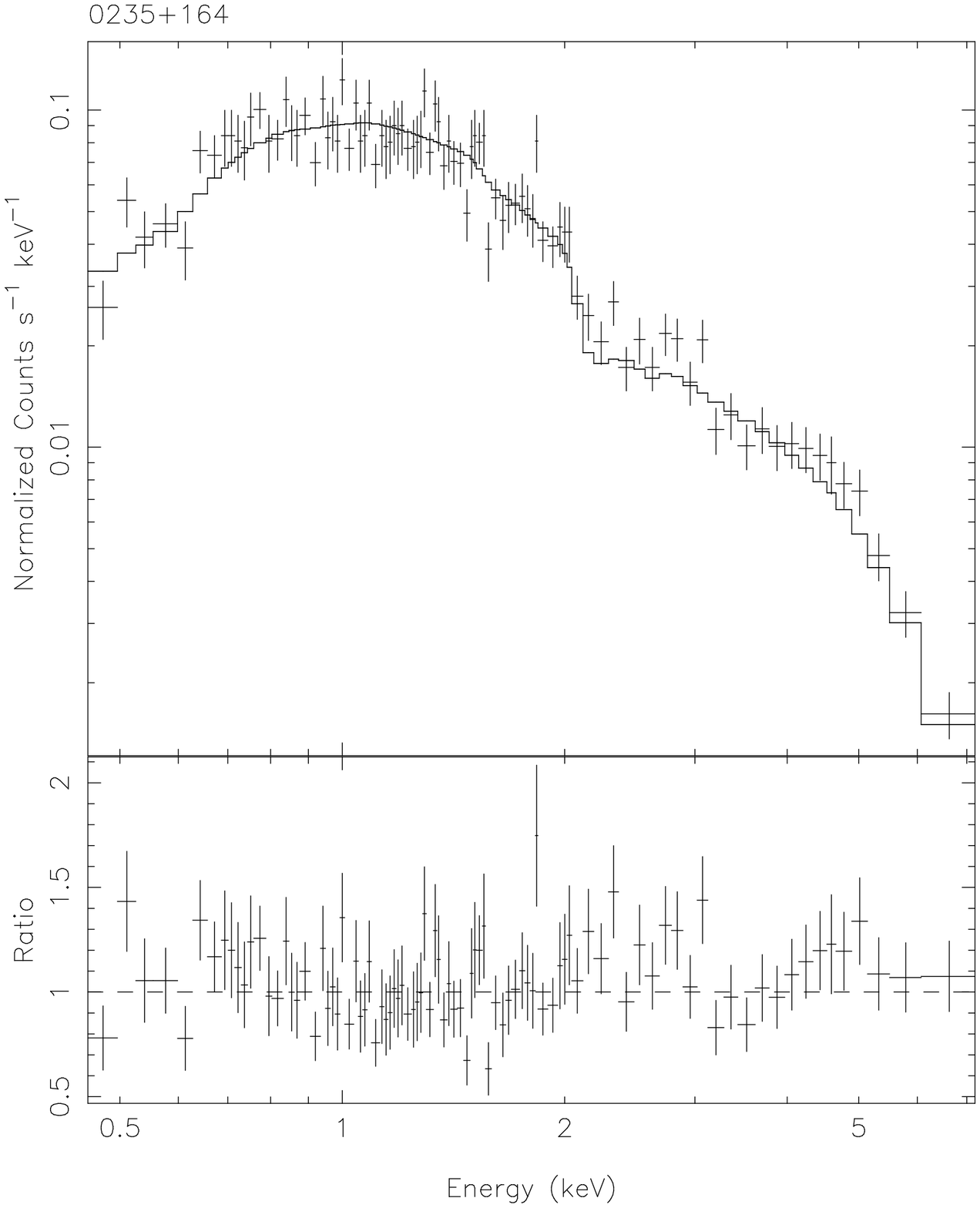}
\caption{$Chandra$ ACIS-S spectrum of the quasar AO 0235+164. The
{\it left panel} shows the power-law fit to the $2-8$ keV region of the
spectrum, extrapolated to 0.4 keV while only accounting for Galactic
ISM absorption. The {\it right panel} shows the same power-law fit
extrapolated to 0.4 kev but including the effects of both Galactic 
ISM absorption and DLA absorption. Details are reported in the 
text and tables.}
\end{figure}

\clearpage

\begin{figure}
\plottwo{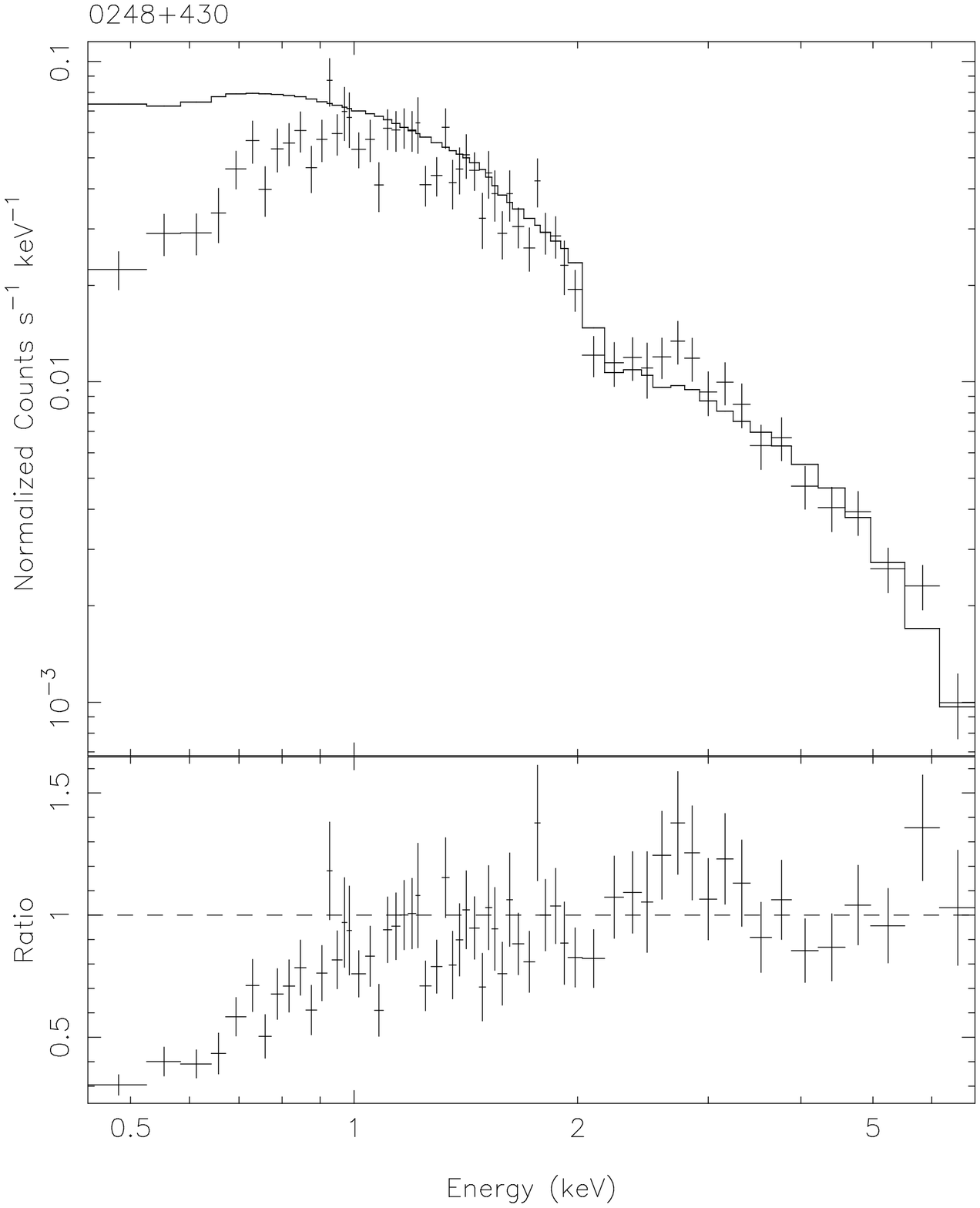}{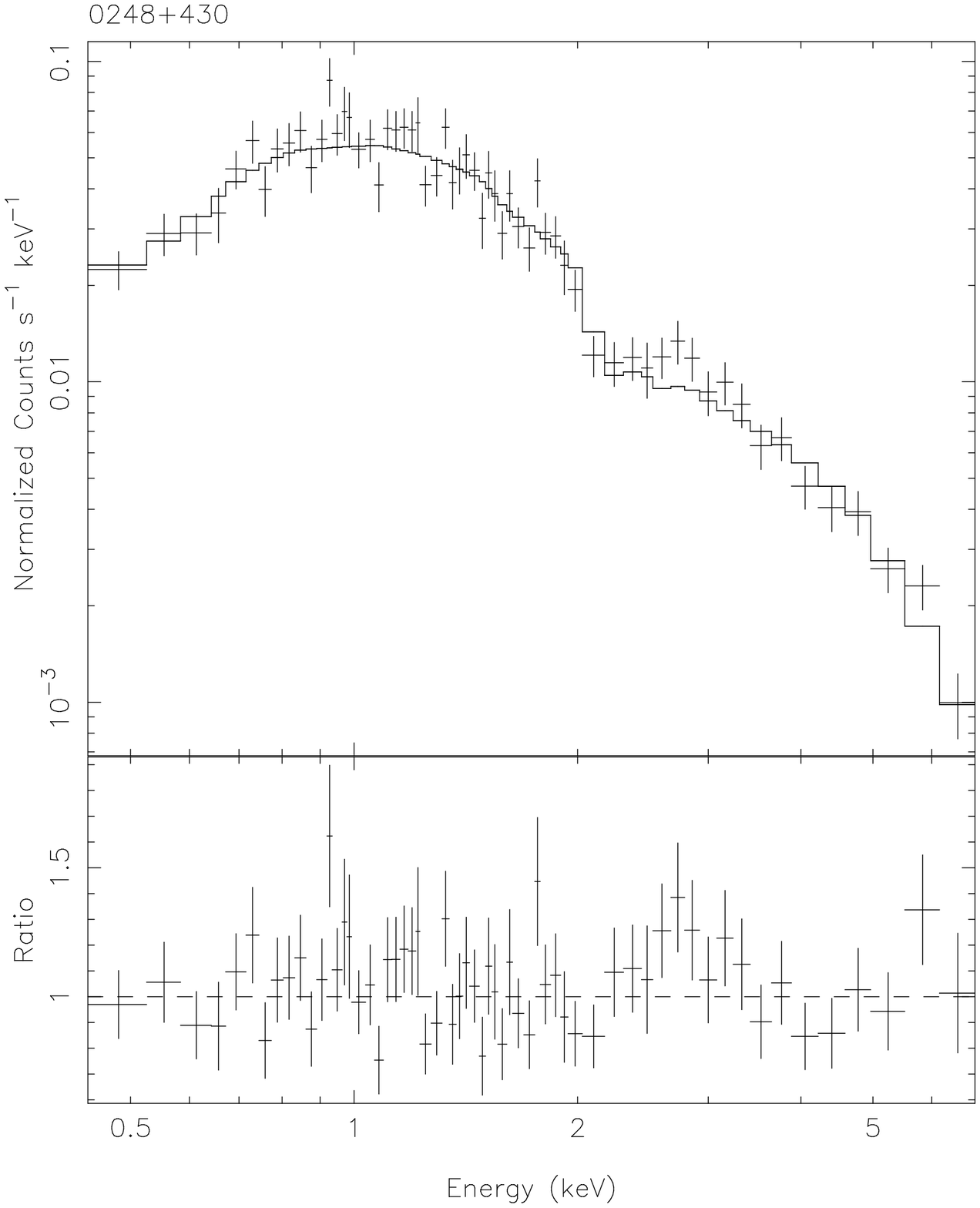}
\caption{$Chandra$ ACIS-S spectrum of the quasar S4 0248+430. The
{\it left panel} shows the power-law fit to the $2-8$ keV region of the
spectrum, extrapolated to 0.4 keV while only accounting for Galactic
ISM absorption. The {\it right panel} shows the same power-law fit
extrapolated to 0.4 kev but including the effects of both Galactic 
ISM absorption and DLA absorption. Details are reported in the 
text and tables.}
\end{figure}

\clearpage

\begin{figure}
\plottwo{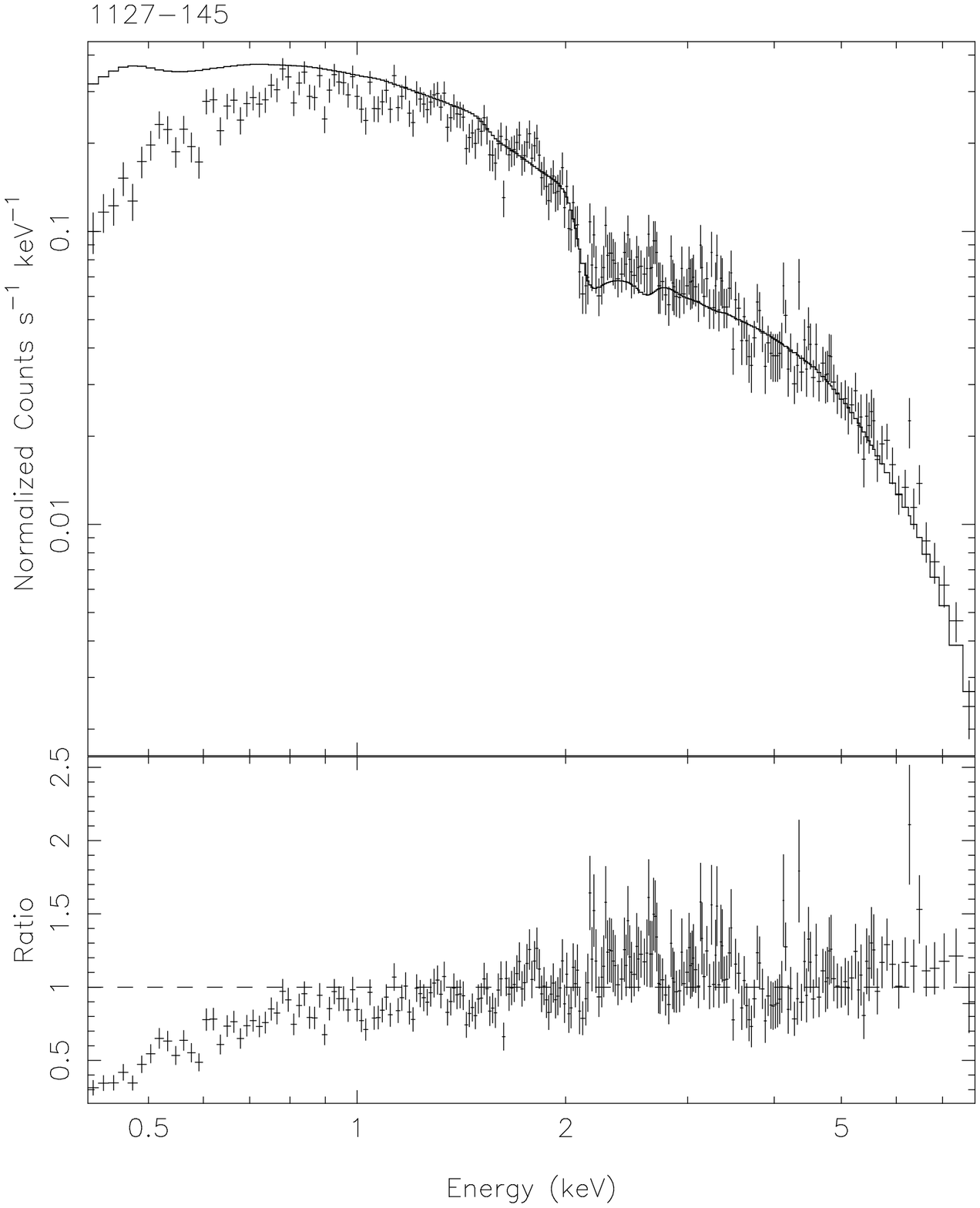}{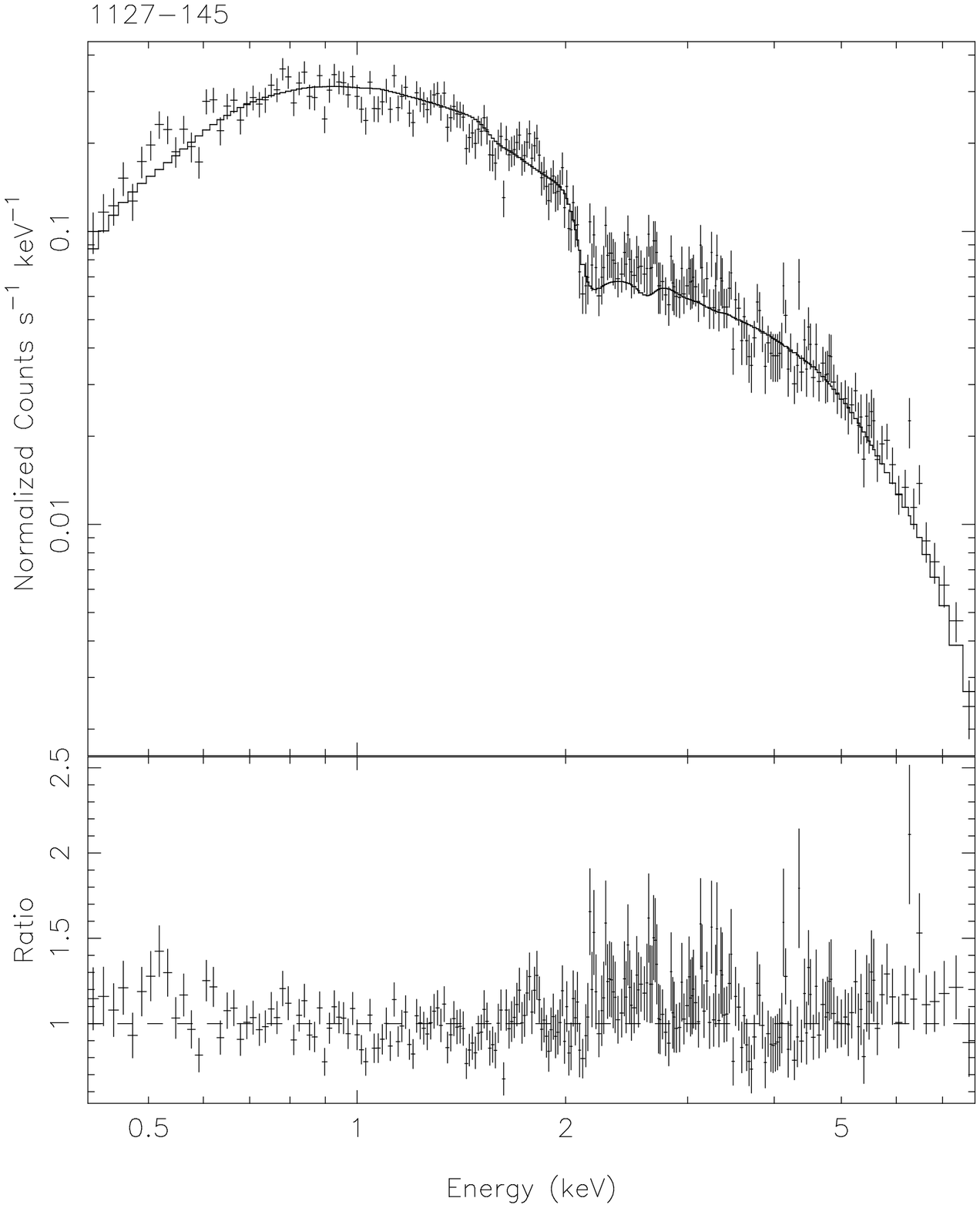}
\caption{$Chandra$ ACIS-S spectrum of the quasar PKS 1127$-$145. The
{\it left panel} shows the power-law fit to the $2-8$ keV region of the
spectrum, extrapolated to 0.4 keV while only accounting for Galactic
ISM absorption. The {\it right panel} shows the same power-law fit
extrapolated to 0.4 kev but including the effects of both Galactic 
ISM absorption and DLA absorption. Note that fixing $N_{HI}$ in the DLA 
system to the value derived from the fit to the DLA line in the $HST$ UV 
spectrum and assuming zero DLA metallicity causes us to under-predict the flux
in the x-ray spectrum in the $0.4 < E < 1$ keV range. 
Details are reported in the text and tables.}
\end{figure}

\clearpage

\begin{figure}
\plotone{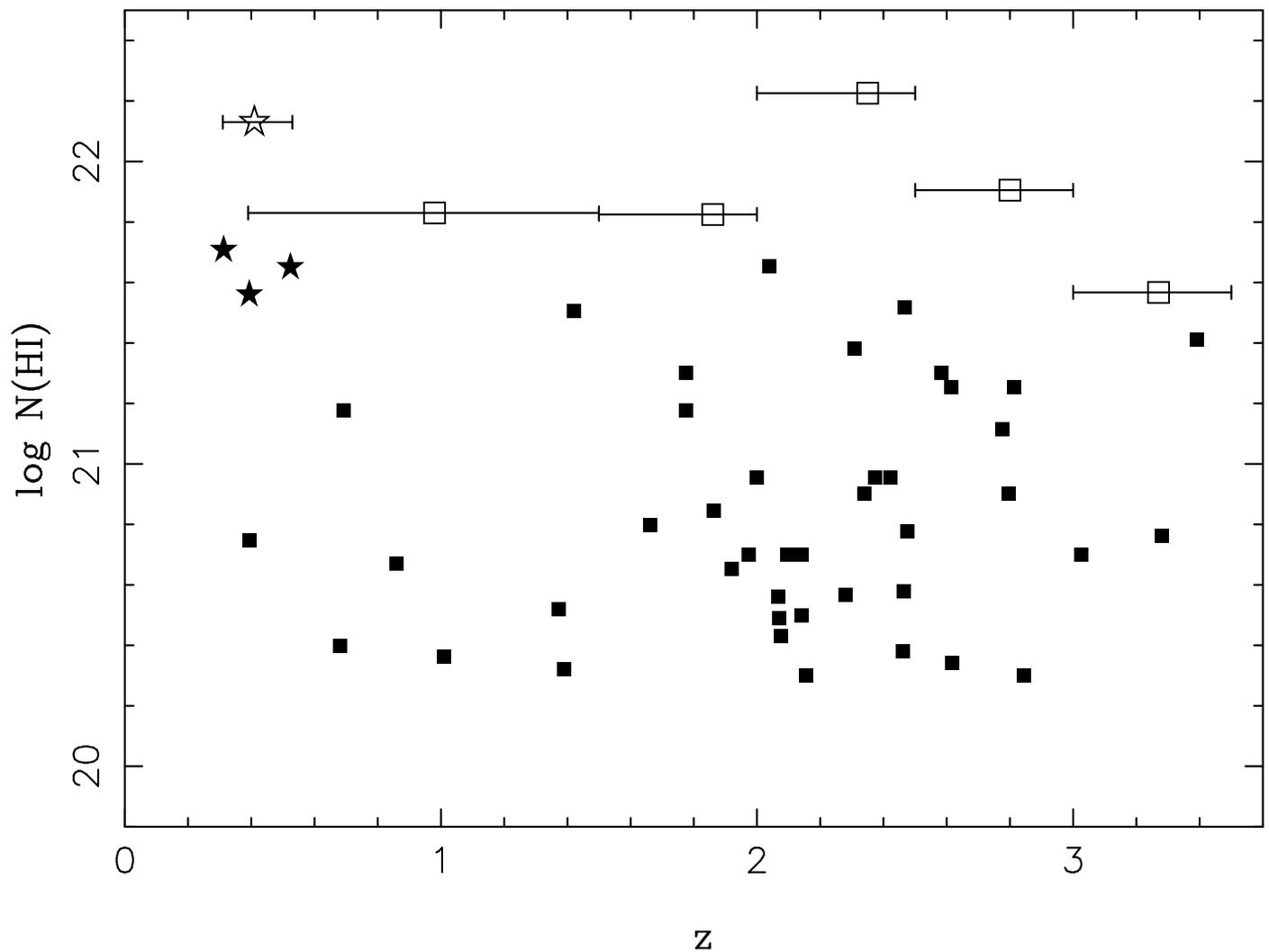}
\caption{Individual HI column densities at the observed
redshifts and summed HI column densities in various redshift
intervals.  The filled star symbols represent the three low-redshift
DLA systems studied here; the filled square symbols represent
the individual higher-redshift DLA systems studied by Pettini et
al. (1999), Prochaska \& Wolfe (1999), Molaro et al. (2000), and
 Ge, Bechtold, \& Kulkarni (2001). Note that some of the data used in 
Pettini et al. (1999) were below the DLA threshold column density 
(not shown here), but these points have little affect on the 
column-density-weighted metallicity results. The unfilled symbols 
are the summed HI column densities in the designated horizontal 
redshift bins.}
\end{figure}

\clearpage

\begin{figure}
\epsscale{.60}
\plotone{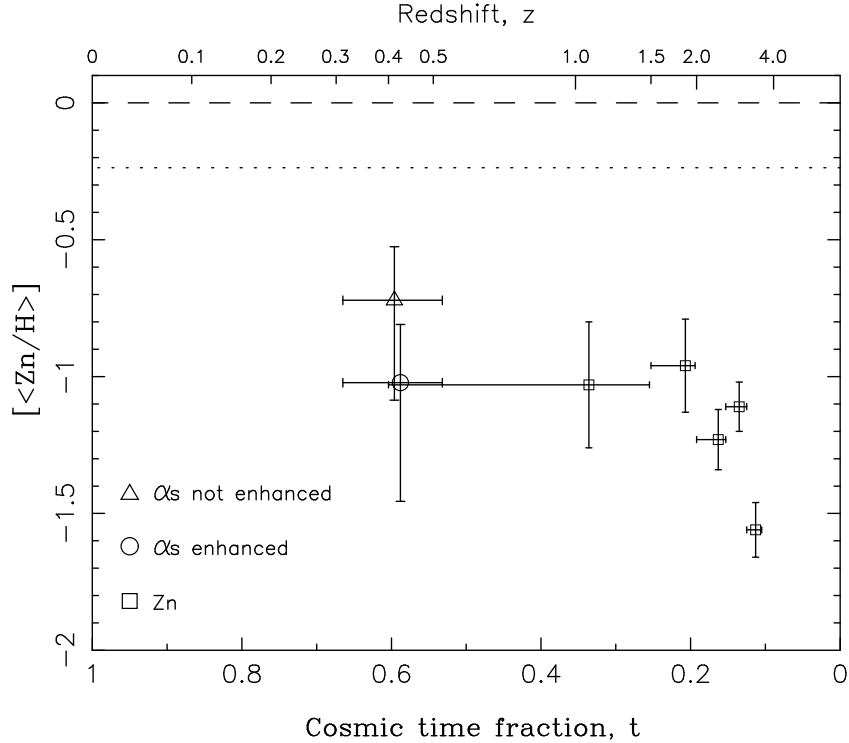}
\caption{Log of the column-density-weighted cosmic
neutral-gas-phase metallicity derived from DLA systems versus cosmic
time (bottom horizontal axis) and redshift (top horizontal axis) for
a cosmology with $H_0=65$ km s$^{-1}$, $q_0=0.5$, and $\Lambda=0$.
The low-redshift metallicity determination was derived from
{\it XSPEC} analysis of the $Chandra$ ACIS-S x-ray spectra of three
quasars containing DLA systems (Table 1 and 2).  The horizontal
``error bars'' represent the redshift intervals covered (e.g.,
$0.313<z_{abs}<0.524$ for the low-redshift point).  The high-redshift
metallicity determinations shown as open squares are those for zinc
(a non-$\alpha$-group element) as reported by Pettini et al. (1999).
The two low-redshift metallicity determinations shown here (which
are slightly displaced in redshift so that they can be distinguished
in the figure) are those derived under the assumption that Galactic
($z=0$) ISM gas absorption follows the metallicity model recommended
by Wilms et al. (2001). Other assumptions are discussed in \S2 of
the text.  The point plotted with an open triangle was derived
under the assumption that the DLA gas has solar ratio metals as
specified by Feldman (1992). The low-redshift point plotted with an
open circle was derived under the assumption that DLA $\alpha$-group
metals where enhanced by a factor of 2.5 over solar ratio metals. By
considering these two cases ($\alpha$-group elements not enhanced
and $\alpha$-group elements enhanced), a fair comparison can be made
with the higher redshift metallicity determinations based on zinc.
The horizontal dashed line indicates solar metallicity; the horizontal
dotted line indicates Galactic ISM gas metallicity.}
\end{figure}


\begin{figure}
\epsscale{.60}
\plotone{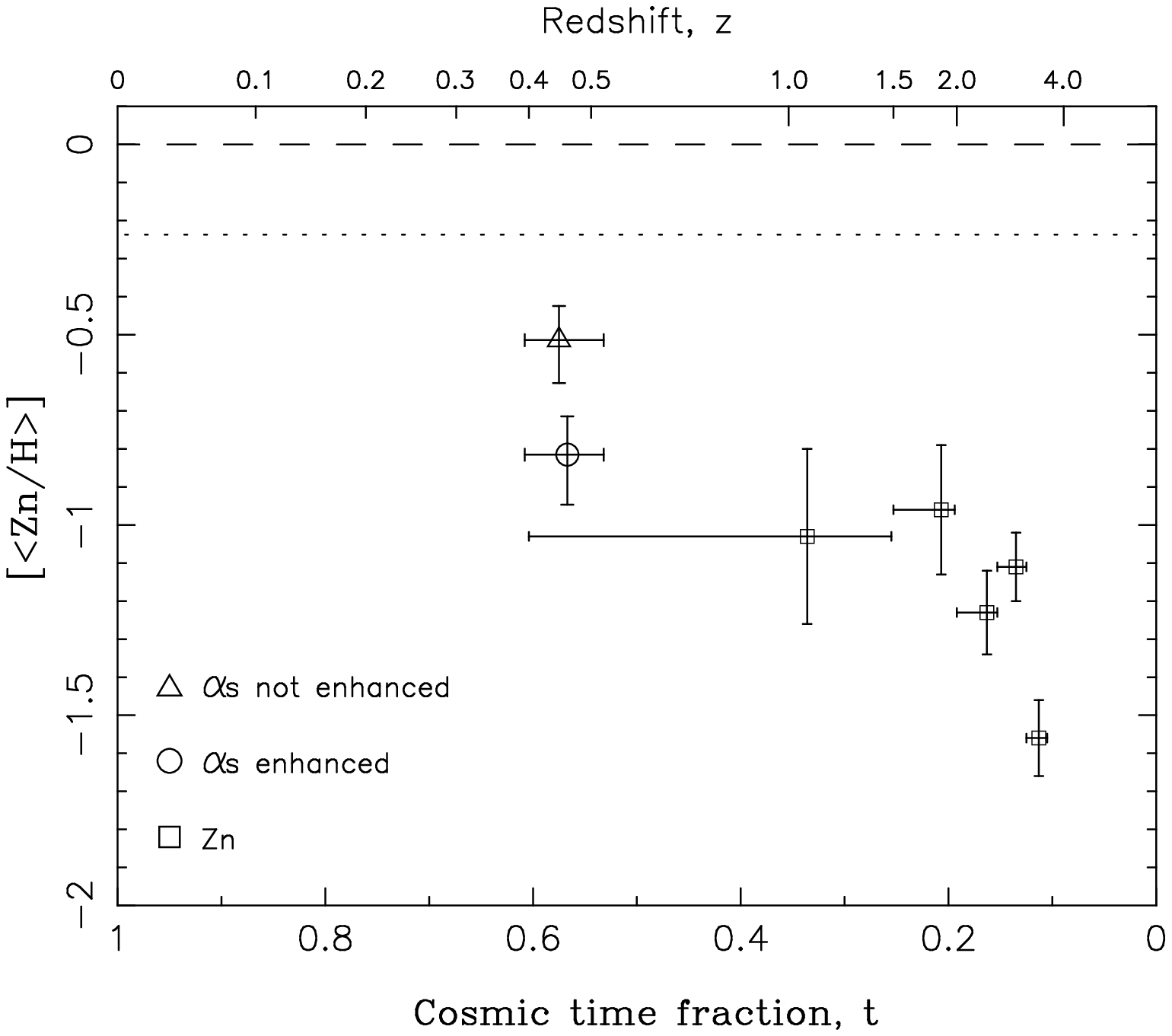}
\caption{ Same as Figure 5, but with the zero
metallicity results on the PKS 1127$-$145 DLA absorber removed. 
The possible justification for removing this DLA absorber from 
the analysis is discussed in \S2 and \S4.1.3.}
\end{figure}

\begin{thebibliography}{}

\bibitem{} Allen, R. 2002, in {\it Seeing Through the Dust -  The
Detection of HI and the Exploration of the ISM in Galaxies}, eds. A.R.
Taylor, T.L. Landecker, and A.G. Willis (ASP Conference Series, in
press), astro-ph/0205168
      
\bibitem{626} Bechtold, J., Siemiginowska, A., Aldcroft, T., Elvis,
M., \& Dobrzycki, A. 2001, ApJ, 562, 133

\bibitem{629} Briggs, F. H. 1983, ApJ, 274, 86

\bibitem{631} Burbidge, E. M., Beaver, E. A., Cohen, R. D.,
Junkkarinen, V. T., \& Lyons, R. W. 1996, AJ, 112, 2533

\bibitem{635} Carilli, C., Lane, W., de Bruyn, A., Braun, R., \&
Miley, G.  1996, AJ, 111, 1830

\bibitem{638} Cohen, R. D., Burbidge, E. M., Junkkarinen, V. T.,
Lyons, R. W., \& Madejski, G. 1999, BAAS, 31, 942

\bibitem{642} Feldman, U. 1992, Phys. Scr., 46, 202

\bibitem{} Ge, J., Bechtold, J., \& Kulkarni, V. 2001, ApJ, 547, L1

\bibitem{644} Hartmann, D., \& Burton, W. B. 1997, {it Atlas of
Galactic Neutral Hydrogen} (Cambridge University Press)

\bibitem{647} Lane, W., Smette, A., Briggs, F., Rao, S., Turnshek,
D., \& Meylan, G. 1998, AJ, 116, 26

\bibitem{650} Lane, W., \& Briggs, F. 2001, ApJ, 561, 27

\bibitem{} Le Brun, V., Bergeron, J., Boiss\'e, P., \& Deharveng, J. M.
1997, A\&A, 321, 733

\bibitem{652} Levshakov, S., Dessauges-Zavadsky, M., D'Odorico, S.,
\& Molaro, P. 2002, ApJ, 565, 696

\bibitem{655} Madejski, G. 1994, ApJ, 432, 554

\bibitem{657} Meyer, D., Jura, M., Hawkins, I., \& Cardelli, J.
1994, ApJ, 437, 59

\bibitem{660} Meyer, D., Jura, M., \& Cardelli, J. 1998, ApJ, 493,
222

\bibitem{} Molaro, P., Bonifacio, P., Centurion, M., D'Odorico, S.,
 Vladilo, G., Santin, P., \& Di Marcantonio, P. 2000, ApJ, 541, 54

\bibitem{662} Murphy, E. M., Lockman, F. J., Laor, A., \& Elvis,
M. 1996, ApJS, 105, 369

\bibitem{665} Nestor, D., Rao, S., Turnshek, D., Monier, E.,  Lane,
W., \& Bergeron, J. 2002, in {\it  Extragalactic Gas at Low Redshift},
eds. J. Mulchaey and J. Stocke, ASP Conference Series, 254, 34

\bibitem{670} Ptak, A. 2001, AAS Mtg, 199, \#145.08

\bibitem{672} Petitjean, P., Srianand, R., \& Ledoux, C.  2002, 332,
383

\bibitem{675} Pettini, M., et al. 1999, ApJ, 510, 576

\bibitem{677} Prochaska, J., \& Wolfe, A. 1999, ApJS, 121, 369

\bibitem{679} Prochaska, J., X., Naumov, S. O., Carney, B. W.,
McWilliam, A., \& Wolfe, A. M. 2000, AJ, 120, 2513

\bibitem{682} Rao, S., \& Turnshek, D. 2000, ApJS, 130, 1 (RT2000)

\bibitem{684} Rao, S., Nestor, D., Turnshek, D., Lane, W., Monier,
E., \& Bergeron, J. 2003, ApJ, submitted, astro-ph/0211297

\bibitem{687} Roberts, M. S., Brown, R. L., Brundage, W. D., \&
Rots, A. H. 1976, AJ, 81, 293

\bibitem{690} Roth, K., \& Blades, J. C. 1995, ApJL, 445, 95

\bibitem{692} Sargent, W., \& Steidel, C. 1990, ApJL, 359, 37

\bibitem{694} Sembach, K., Steidel, C., Macke, R., \& Meyer, D.
1995, ApJL, 445, 27

\bibitem{697} Siemiginowska, A., Bechtold, J., Aldcroft, T. L.,
Elvis, M., Harris, D. E., \& Dobrzycki, A. 2002, ApJ,  570, 543

\bibitem{701} Turnshek, D., Rao, S., Nestor, D., Lane, W., Monier,
E., Bergeron, J., \& Smette, A. 2001, ApJ, 553, 288

\bibitem{704} Turnshek, D., Rao, S., \& Nestor, D. 2002, in {\it
Extragalactic Gas at Low Redshift}, eds. J. Mulchaey and J. Stocke,
ASP Conference Series, 254, 42

\bibitem{708} Vladilo, G., Centurion, M., Bonifacio, P., \& Howk,
J. C. 2001, ApJ, 557, 1007

\bibitem{711} Wilms, J., Allen, A., \& McCray, R. 2000, 542, 914

\bibitem{713} Wolfe, A. M., Briggs, F. H., \&  Davis, M. M. 1982,
ApJ, 259, 495

\bibitem{716} Womble, D., Junkkarinen, V., Cohen, R., \& Burbidge,
M.  1990, 100, 1785

\end{thebibliography}
\end{document}